\documentclass{conm-p-l}

\theoremstyle{definition}

\theoremstyle{remark}

\numberwithin{equation}{section}
  \copyrightinfo{2006}%            % copyright year
    {American Mathematical Society}% copyright holder
\begin{document}
\title{Noncommutativity from Canonical and Noncanonical Structures}
\author{Marcos Rosenbaum}
\address{Instituto de Ciencias Nucleares, UNAM, A. Postal
70-543,M\'exico D.F.,  M\'exico.} \email{mrosen@nucleares.unam.mx}
\thanks{The first author was supported in part by CONACyT Grant \#UA7899-F}
\author{J. David Vergara}
\address{Instituto de Ciencias Nucleares, UNAM,
A. Postal 70-543,M\'exico D.F.,  M\'exico.}
\email{vergara@nucleares.unam.mx}
\thanks{The second author was supported in part by grants:
DGAPA-IN104503, CONACyT 47211-F}
\author{L. Rom\'an Ju\'arez}
\address{Instituto de Ciencias Nucleares, UNAM,
A. Postal 70-543,M\'exico D.F.,  M\'exico.}
\email{lromanjs@gmail.com}
\thanks{The third author was partially supported by a SNI fellowship.}
%    General info
\date{March 9, 2006.}
\dedicatory{This paper is dedicated to J. Plebanski.}

\begin{abstract}
Using arbitrary symplectic structures and parametrization invariant
actions, we develop a formalism, based on Dirac's quantization
procedure, that allows us to consider theories with both space-space
as well as space-time noncommutativity. Because the formalism has as
a starting point an action, the procedure admits quantizing the
theory either by obtaining the quantum evolution equations or by
using the path integral techniques. For both approaches we only need
to select a complete basis of commutative observables. We show that
for certain choices of the potentials that generate a given
symplectic structure, the phase of the quantum transition function
between the admissible bases corresponds to a linear canonical
transformation, by means of which the actions associated to each of
these bases may be related and hence lead to equivalent
quantizations. There are however other potentials that result in
actions which can not be related to the previous ones by canonical
transformations, and for which the fixed end-points, in terms of the
admissible bases, can only be realized by means of a Darboux map. In
such cases the original arbitrary symplectic structure is reduced to
its canonical form and therefore each of these actions results in a
different quantum theory. One interesting feature of the formalism
here discussed is that it can be introduced both at the levels of
particle systems as well as of field theory.
\end{abstract}
\maketitle
\section{Introduction}
\setcounter{equation}{0} In recent years, space-time
noncommutativity has become the subject of increasing interest. In
field theory stimulated by some results in low energy string theory,
and in quantum mechanics because it is in the context of this
formalism that space-time noncommutativity is more naturally
understood in terms of space and time operators acting on a Hilbert
space and also, because quantum mechanics viewed as a minisuperspace
reduction of field theory, could reasonably be expected to
provided further insight into how quantum mechanical
noncommutativity reflects itself in field theory. Some of the more
relevant work related to the approach here considered may be found
in \cite{Doplicher}, \cite{Sibold:1}, \cite{Fujikawa},
\cite{Balachandran:1},
\cite{Pinzul}, \cite{Heslop}, \cite{Banerjee}.\\
An interesting idea that allows us to consider in a full setting the
space-time noncommutativity in the context of particle mechanics, is
to use the concept of parametrization invariance \cite{Pinzul},
\cite{Banerjee}. In this way the time is taken as an extra canonical
variable of the system and it is then easy to introduce a
non-canonical structure in this extended phase- space. The usual way
to study the parametrization invariance of a system is by using the
Dirac method of canonical analysis. Because not all the momenta are
independent due to the invariance under parametrizations, this
approach requires that a constraint on the system be introduced. For
a parametrized particle, this constraint is at the classical level
the Hamilton-Jacobi equation and at the quantum level the
Schrodinger equation. So the Dirac method associates to the symmetry
of parametrizations the classical or quantum evolution equations
\cite{Henneaux}.

Here we want to generalize the above mentioned procedure in order to
be able to consider noncommutative theories at the quantum level
resulting both from canonical and non-canonical structures. The
noncommutativity will then appear as a consequence of the existence
of second class constraints, and the implementation of these
constraints in terms of Dirac brackets. The interesting point of the
procedure is that on the one hand we get the classical and quantum
evolution equations for the noncommutative systems and on the other
hand we also obtain a classical action that can be quantized using
the path integral formalism. Furthermore, the analysis is not
restricted to noncommutative theories with constant deformation
parameters, since the procedure naturally incorporates arbitrary
canonical potentials. Another interesting property of the method is
that it can be naturally extended to field theory.

Our starting point is to consider a parametrization invariant
system. This means that if the system is not naturally invariant
under parametrizations we promote the original parameters of the
theory, for example the time in the case of particle dynamics, to
the level of canonical variables. The second step is to perform the
canonical analysis of this theory. One point that we must be careful
with is that, since we add new variables to the system,
 we have to introduce constraints associated to the parametrization invariance
 symmetry of the theory in order that the number of degrees of freedom are preserved. The
third step is to introduce an arbitrary canonical potential that
allows us to realize the required noncommutativity. The next step is
to show that under the Dirac brackets the first class constraint (or
constraints) generate the symmetry. This means that
 we will probably need to modify the constraints. At this point, if we
have several constraints, we need to check that the algebra of these
first class constraints closes. Once we finish this procedure we
obtain the quantum evolution equations for our system.
Alternatively, we can introduce the canonical potential in the
action and select an appropriate basis in order to quantize the
system using the path integral formalism. For certain choices of the
potentials that generate a given symplectic structure, the phase of
the quantum transition function between the admissible bases
corresponds to a linear canonical transformation, by means of which
the actions associated to each of these bases may be related and
hence lead to equivalent quantizations. We must stress that in
contradistinction to the case when time plays the role of a
parameter, the canonical transformation here is implemented in
an extended phase space,
where the time and its conjugate momentum are included.

With the purpose of examining all the above mentioned facets of the
space-time noncommutativity, our presentation has been structured as
follows: In Section 2 we consider the canonical formalism of
parametrization invariant systems. In Section 3 we introduce an
arbitrary symplectic structure in the action, and after the
canonical analysis we construct the Dirac brackets associated to the
theory and also obtain the action for the reduced system.  In
Section 4, we quantize the theory using different bases, and using
both path integral methods and the quantum evolution equations. We
conclude the paper with some remarks and possible extensions.

\section{Parametrization invariant systems}
\setcounter{equation}{0}
 We begin here by reviewing the essentials of the canonical analysis
of parametrized systems following the approach in \cite{Henneaux}.
To this end, consider the action for a particle in a
 $N$-dimensional configuration space, in an
arbitrary potential:
\begin{equation}\label{act0}
    S = \int\limits_{t_1 }^{t_2 } {dt\left( {\frac{1}{2}m\left(
{\frac{{dq^i }}{{dt}}} \right)^2  - V\left( {q^i },t \right)}
\right)},
\end{equation}
where $i=1,\dots, N$. In this action the time $t$ plays the role of
a parameter in the theory. To study the non-commutativity of the
space and time it is more convenient to consider the time as another
coordinate of our theory, i.e. we extend our configuration space
with one extra dimension $t=q^0$. To do this, we parametrize the
action by introducing a new parameter $\tau$ and assume that the
coordinates $q^i(t)$ are scalars under this parametrization, i.e.,
\begin{equation}\label{pareme}
    \begin{split}
     t&\to \tau\\
        q^i \left( t \right) &\to q^i \left( \tau  \right)
\end{split}
\end{equation}
The action (\ref{act0}) takes the form
\begin{equation}\label{actionpare}
S = \int\limits_{\tau _1 }^{\tau _2 } {d\tau \left(
{\frac{1}{2}m\left( {\frac{{dq^i }}{{d\tau }}} \right)^2 \left(
{\frac{{d\tau }}{{dt}}} \right) - V\left( {q^i,t } \right)\left(
{\frac{{dt}}{{d\tau }}} \right)} \right)},
\end{equation}
where $t=q^0$ now plays the role of a new coordinate in the theory.
Making the identifications $ \dot q^i  \equiv \left( {\frac{{dq^i
}}{{d\tau }}} \right)$ and $ \dot q^0 \equiv \frac{{dt}}{{d\tau }}$,
we can rewrite (\ref{actionpare}) in the form
\begin{equation}\label{actionpare1}
S = \int\limits_{\tau _1 }^{\tau _2 } {d\tau } \left(
{\frac{1}{2}m\frac{{(\dot q^i)^2 }}{{\dot q^0}} - V(q^i,q^0 )\dot
q^0} \right).
\end{equation}

In Hamiltonian form the action (\ref{actionpare1}) reads
\begin{equation}\label{actionhamp}
S = \int\limits_{\tau _1 }^{\tau _2 } {d\tau } \left( {p_0} \dot q^0
+ p_i \dot q^i  - \lambda \varphi \right ),
\end{equation}
where  $\varphi=p_0  + H\approx 0$ is the first class primary
constraint associated to the symmetry under parametrizations (which
needs to be included in (\ref{actionhamp}) in order to account for
the fact that by introducing a new variable in the theory,
restrictions must be added to the physical evolution of the system
that indicate that the $N+1$ new coordinates are not all
independent),
 $H$ is the canonical Hamiltonian of the action (\ref{act0}), and $\lambda(\tau)$
  is a Lagrange multiplier.
The action (\ref{actionhamp}) is invariant up to a total derivative
under the transformations generated by the constraint $\varphi$,
given by
\begin{equation}\label{varia}
\delta q_0  = \left\{ {q_0 ,\varepsilon \varphi } \right\}, \ \
\delta p_0  = \left\{ {p_0 ,\varepsilon \varphi } \right\}, \ \
\delta p_i  = \left\{ {p_i ,\varepsilon \varphi } \right\}, \ \
 \delta q^i  = \left\{ {q^i ,\varepsilon \varphi } \right\} \ \
 \delta\lambda=\dot \varepsilon,
\end{equation}
where the variation of the Lagrange multiplier is imposed in such
way that
when varying the action it should vanish up to a boundary term.\\

Following Dirac \cite{Dirac}, we propose that at the quantum level
the physical sates of the theory are invariant under the above
transformations, i.e.,
\begin{equation}\label{quantuminv}
e^{i\varepsilon \hat \varphi } \left| \psi  \right\rangle _P  =
\left| \psi  \right\rangle _P.
\end{equation}
So in infinitesimal form we get
\begin{equation}\label{quantuminv1}
\hat \varphi \left| \psi  \right\rangle _P  = 0.
\end{equation}
We thus see that the constraint leads to a supplementary condition
on the physical states, and is another way to reduce the quantum
theory to its physical sector
without imposing a gauge condition.\\

Now if we consider the configuration representation with basis
$|q^0,q^i\rangle$, equation (\ref{quantuminv1}) yields,
\begin{equation}
\hat \varphi \left| \psi  \right\rangle _P  = 0 \Rightarrow \left( {
- i\hbar\frac{\partial }{{\partial t}} - \frac{\hbar^2}{{2m}}\nabla
^2 + V(q^i,t )} \right)\psi (q^i ,t) = 0,
\end{equation}
where we have identified $t=q^0$. We therefore obtain the
Schrodinger equation as a result of imposing at the quantum level
the classical invariance under parametrizations of the theory.\\
In the following section we shall apply the same procedure to the
case of arbitrary symplectic structures.

\section{Non-commutativity and Dirac Brackets}
\setcounter{equation}{0}
 Let $z^a=\left( {q^0,q^i ,p_0 ,p_i
}\right)$,
 with $a = 1,...,2N + 2$, denote the $2N+2$ phase-space variables of
a parametrized system in the Hamiltonian formulation. In this case
we don't have a second order action to begin with as in
(\ref{act0}). We can however consider a general first order action,
equivalent to (\ref{actionhamp}), given by
\begin{equation}\label{nonaction1}
S = \int\limits_{\tau _1 }^{\tau _2 } {d\tau } \left( {A_a (z)\dot
z^a  - \lambda \varphi(z) } \right),
\end{equation}
where $A_a(z)$ is a vector potential which we shall use to generate
an arbitrary symplectic structure associated to the Poisson brackets
in the Hamiltonian formulation.

Applying the Dirac's method for constrained systems, we have from
(\ref{nonaction1}) that the corresponding canonical Hamiltonian is
given by
\begin{equation}\label{canicalham}
H_c  = \lambda \varphi(z),
\end{equation}
and the canonical momenta  lead to the set of primary constraints,
\begin{equation}\label{pricons}
\chi _a  = p_{z a}  - A_a \left( z \right).
\end{equation}
Consequently, the total Hamiltonian for this theory is
\begin{equation}\label{HT}
H_T  = \lambda \varphi + \mu ^a \chi _a.
\end{equation}
Moreover, from the evolution of the constraints we obtain the
following consistency conditions
\begin{equation}\label{evocons}
\dot \chi _a  = \left\{ {p_{z a}  - A_a \left( z \right),H_T }
\right\} =  - \lambda \frac{{\partial \phi }}{{\partial z^a }} + \mu
^b \omega _{ab}  \approx 0,
\end{equation}
where
\begin{equation}\label{symstr}
\omega _{ab} : = \partial _a A_b  - \partial _b A_a  = \{ \chi _a
,\chi _b \}.
\end{equation}
This antisymmetric matrix will play the role of the symplectic
structure of the theory. Assuming further that $\omega _{ab}$ is
invertible so all the Lagrange's multipliers $\mu^a$ in
(\ref{evocons}) can be determined, it then follows from
(\ref{symstr}) that the constraints $\chi_a$ are second class. Note
that in the case where the symplectic structure is degenerate, at
least one of the $\chi_a$'s will be first class, but in this case
the number of degrees of freedom of the generalized theory will not
correspond to the degrees of freedom of the original theory. Hence
in what follows we will
assume that all the constraints $\chi_a$ are second class.\\

Now, in order to impose these constraints as strong conditions when
quantizing, we construct the associated Dirac brackets which are
given by
\begin{equation}\label{diracbra}
\left\{ {A,B} \right\}^*  = \left\{ {A,B} \right\} - \left\{ {A,\chi
_a } \right\}\omega ^{ab} \left\{ {\chi _b ,B} \right\},
\end{equation}
where $\omega ^{ab}$, is the inverse matrix of $\omega _{ab}$.
Computing the Dirac's brackets of the coordinates with the above
expression we obtain
\begin{equation}\label{diracbraz}
\left\{ {z^a ,z^b } \right\}^*  = \omega ^{ab}.
\end{equation}
Thus, quantizing a theory constrained by symmetries under
parametrization results in the noncommutativity of the quantum
operators corresponding to the phase space coordinates:
\begin{equation}\label{commutz}
\left[ {\hat z}^a ,{\hat z}^b  \right]  = i\hbar \omega ^{ab}.
\end{equation}
The simplest case corresponds to the usual Heisenberg algebra of
ordinary Quantum Mechanics, for which the inverse matrix of the
canonical symplectic structure takes the form
\begin{equation}\label{omega}
    J^{ab}:=\omega^{ab}|_{\theta=0}=\left(%
\begin{array}{cc}
  0 & \mathbb{I} \\
  \mathbb{-I} & 0 \\
\end{array}%
\right).
\end{equation}

\section{Non-commutative Quantum Mechanics}
In the previous section we have considered a general procedure for
quantizing a theory with an arbitrary symplectic structure. One
interesting feature of this formalism is that by including time as a
canonical variable allows us to consider also noncommutativity
between the time and the spatial coordinates. Now, given such a
symplectic structure we can quantize either by using the Dirac's
procedure where the first class constraints act as operators on the
physical states, imposing supplementary conditions on them, and the
Dirac brackets of the second class constraints are replaced by
commutators,  or, alternatively, we can also quantize by first
evaluating the generating potentials of the symplectic structure and
then applying path integral methods in order to derive
the Feynman propagators.\\

It should be noted, however, that for a given symplectic structure
the solution for the potentials $A_a$ is not unique, although all
the possible resulting actions and resulting classical theories are
related by  canonical transformations.\\
Furthermore, in the Dirac quantization the commutators
(\ref{commutz}) of the generators of the extended Heisenberg algebra
define the possible complete sets of commuting observables of the
theory and the correlative admissible bases (labeled by the
eigenvalues of these sets). For each of these admissible bases, we
obtain a realization of the Heisenberg algebra and of the subsidiary
condition (\ref{quantuminv1}) and, correspondingly in the path
integral formalism, the Feynman propagators derived from the
transition functions in each of these bases. This means that in the
path integral calculation of a transition function, the only
admissible actions are those for which the fixed end-points in a
variational principle are the same as the dynamical variables
labeling the basis used for the
evaluation of the transition function.\\

Note finally that there are also actions originating from solutions
of (\ref{symstr}) for which no fixed end-points, corresponding to
one of the admissible bases in the Dirac quantization exits.
However, can be defined using a Darboux map.  This map, involves
introducing new dynamical variables in terms of linear combinations
of the original ones and, consequently implies a change in the
initial symplectic structure to a canonical one. Compatible,
although non-equivalent, path integral and Dirac quantizations
result from promoting to the rank of operators these new variables,
which will satisfy the Heisenberg algebra of ordinary quantum
mechanics. So in these cases the deformation of the symplectic
structure at the classical level is reflected at the quantum level
in a deformed Hamiltonian while the
standard Heisenberg algebra of the usual quantum mechanics is preserved. \\

To further illustrate the above observations, we next consider some
examples of quantum noncommutativity schemes in the context of both
the Dirac and path integral formalisms. For analytical simplicity we
assume a 1+1 space-time, generalization to higher order dimensions
is fairly straightforward.

\subsection{Space-time noncommutativity}
Let us consider first the case where the Dirac brackets
(\ref{diracbraz}) determine a symplectic structure of the form
\begin{equation}\label{oab1}
\omega ^{ab}  = \left( {\begin{array}{*{20}c}
   0 & \theta  & 1 & 0  \\
   { - \theta } & 0 & 0 & 1  \\
   { - 1} & 0 & 0 & 0  \\
   0 & { - 1} & 0 & 0  \\
\end{array} } \right),  \ \ \
\omega _{ab}  = \left( {\begin{array}{*{20}c}
   0 & 0 & { - 1} & 0  \\
   0 & 0 & 0 & { - 1}  \\
   1 & 0 & 0 & \theta   \\
   0 & 1 & { - \theta } & 0  \\
\end{array} } \right).
\end{equation}
Quantizing according to Dirac's prescription by using
(\ref{commutz}) leads to the commutators
\begin{equation}\label{commu}
[\hat t,\hat x] = i\hbar\theta, \ \ [\hat x,\hat p_x] = i\hbar,\ \
[\hat t,\hat p_t] = i\hbar ,\ \ [\hat p_t,\hat p_x] = 0,
\end{equation}
 and, using (\ref{quantuminv1}), to the supplementary condition
\begin{equation}\label{qsup}
\hat\varphi|\psi\rangle=0,
\end{equation}
where $\hat\varphi$ is given by
\begin{equation}\label{cons2}
\hat\varphi= \hat p_t + H(\hat t,\hat x,\hat p_x).
\end{equation}
It is obvious from (\ref{commu}) that for a mechanical Hamiltonian
the sets of complete commuting observables in this case are $\{\hat
x,\hat p_t\}$, $\{\hat t,\hat p_x\}$ and $\{\hat p_t,\hat p_x\}$.
The admissible bases in Hilbert space are then  $\{|x,
p_t\rangle\}$, $\{|t, p_x\rangle\}$ and $\{|p_t, p_x\rangle\}$,
respectively.\\

\subsubsection{Basis $|x,p_t\rangle$}
For the basis $\{\hat x,\hat p_t\}$ the algebra
(\ref{commu}) is realized by
\begin{equation}\label{rea2}
    \hat t \psi(x,p_t)=i\hbar \left(\partial_{p_t}+\theta
    \partial_{x}\right)\psi(x,p_t), \ \ \ \ \hat p_x \psi(x,p_t)=-i\hbar
    \partial_x\psi(x,p_t),
\end{equation}
while the remaining generators of the extended Heisenberg algebra
are just multiplicative quantities. Also projecting on (\ref{qsup})
with $\langle x, p_t |$ and substituting (\ref{rea2}) into
(\ref{cons2}), with  a Hamiltonian of the form
$H=\frac{p_x^2}{2m}+V(x,t)$, yields the subsidiary condition
\begin{equation}\label{schro}
\left(p_t -\frac{\hbar^2}{2m}\partial_x^2 +
V\left(x,i\hbar(\partial_{p_t}+\theta
\partial_x)\right)\right)\psi(x,p_t)=0
\end{equation}
on the wave function $\psi(x,p_t)$.\\
One interesting feature of the Dirac quantization resulting from the
use of this basis is that for a $t$ independent potential, equation
(\ref{schro}) becomes
\begin{equation}\label{schro2}
\left(p_t -\frac{\hbar^2}{2m}\partial_x^2 +
V\left(x\right)\right)\psi(x,p_t)=0.
\end{equation}
 For such a time independent Hamiltonian, (\ref{schro2}) may be interpreted
as an eigenvalue equation, with
 $-p_t$ the energy eigenvalues of the system and $\psi(x,p_t)$
the corresponding eigenvectors.
Note that the energy spectrum of the resulting theory does not have
 any corrections from the noncommutativity of the space-time. A similar result was obtained by
Balachandran, et al \cite{Balachandran:1} by means of a very
different approach.\\

Now, in order to obtain the equivalent quantization by means of path
integrals, we need to compute the transition function $\langle
x(\tau_2), p_t(\tau_2)|x(\tau_1), p_t(\tau_1)\rangle$. For this
purpose we need first to derive the appropriate action function,
which according to our previous observations has to have as fixed
end-points the variables $x, p_t$. This, as implied  by
(\ref{symstr}), requires in turn deriving the proper generating
potentials $A_a(z)$ for the symplectic structure (\ref{oab1}) by
solving the equations,
\begin{equation}\label{paras}
\frac{{\partial A_1 }} {{\partial p_t }} - \frac{{\partial A_3 }}
{{\partial t}} = 1, \ \ \frac{{\partial A_2 }} {{\partial p_x }} -
\frac{{\partial A_4 }} {{\partial x}} = 1, \ \ \frac{{\partial A_4
}} {{\partial p_t }} - \frac{{\partial A_3 }} {{\partial p_x }} =
\theta.
\end{equation}
It is not difficult to verify that the needed solution is
\begin{equation}\label{acb1}
     A_1  = 0,\; A_2  = p_x, \;A_3  = -(t+\theta p_x),\; A_4  = 0.
\end{equation}
In fact, Inserting (\ref{acb1}) in the action (\ref{nonaction1})
results in
\begin{equation}\label{act6}
S_1 = \int\limits_{\tau _1 }^{\tau _2 } {d\tau } \left( {p_x \dot x
-\theta p_x \dot p_t    -t \dot p_t - \lambda \left( {p_t +
H(t,x,p_x)} \right)} \right),
\end{equation}
which indeed has the appropriate variational fixed end-points $x,
p_t$.
With (\ref{act6}) we can now compute the propagator
\begin{equation}\label{propa1}
\langle x(\tau_2), p_t(\tau_2)|x(\tau_1), p_t(\tau_1)\rangle = \int
{\mathcal D}t {\mathcal D}p_t {\mathcal D}x {\mathcal D}p_x
\delta(\chi)\delta(\varphi)\{\varphi,\chi\}^* \exp (\frac{i}{\hbar}
S_1),
\end{equation}
where we have introduced a canonical gauge fixing condition
$\chi=\chi(\tau,t,p_t,x,p_x)$. This gauge must first be a good
canonical gauge in the Dirac's sense, i.e. the Dirac bracket
$\{\varphi,\chi\}^*$ must be invertible and second the gauge must be
consistent with the boundary conditions. Because, we are fixing at
the end points $(x,p_t)$, it is not possible to use the usual gauge
$t=f(\tau)$, we will use instead the gauge condition
\begin{equation}\label{g2}
   \chi= x-f(\tau)\approx 0.
\end{equation}
The Dirac's bracket between this gauge condition and the constraint
is given by
\begin{equation}\label{db1}
\{\varphi, \chi\}^* = -\frac{p_x}{m}+\theta\frac{\partial
V}{\partial t}
\end{equation}
This gauge is a good canonical gauge for $p_x\neq 0$, in which case
the path integral has two different branches, one corresponding to
$p_x>0$ and the other for negative $p_x$. It can also be seen that
this term leads to corrections of first order in $\theta$ which are,
however, proportional to the time dependence of the potential.
Consequently, if we assume that the potential is time independent,
this corrections cancel and we can then integrate (\ref{propa1})
over $t$ to obtain
\begin{equation}\label{propa2}
\begin{split}
\langle x(\tau_2), p_t(\tau_2)|x(\tau_1), p_t(\tau_1)\rangle =\hspace{2in} \\
\int{\mathcal D}x {\mathcal D}p_t {\mathcal D}p_x
\delta(x-f)\delta(\varphi)\delta(\dot p_t) \left(-\frac{p_x}{m}\right)\times \\
\left( \exp \left(\frac{i}{\hbar} \int_{\tau_1}^{\tau_2}d\tau
\left(p_x(\dot x -\theta \dot p_t)-\lambda
\varphi(p_t,p_x,x)\right)\right)\right).
\end{split}
\end{equation}

Note now  that the only dependence on $\theta$ in the above
expression appears multiplying $\dot p_t$, but taking into account
that this term is zero due to the delta functional in the path
integral we do not get nonconmmutative corrections to the
propagator. This is in agreement with our previous results derived
by using the Dirac's quantization.\\

\subsubsection{Basis $|t,p_x \rangle$} Let us next consider the basis $\{\hat t,\hat p_x\}$
in which the operators $\hat x$ and $\hat p_t$ are realized by
\begin{equation}\label{rea1}
    \hat x \psi(t,p_x)=i\hbar \left(\partial_{p_x}-\theta
    \partial_{t}\right)\psi(t,p_x), \ \ \ \ \hat p_t \psi(t,p_x)=-i\hbar
    \partial_t\psi(t,p_x).
\end{equation}
In the Dirac quantization we have that a realization of the
supplementary condition (\ref{quantuminv1}) in this basis results
from projecting with $\langle t, p_x |$ and substituting
(\ref{rea1}) into the first class constraint (\ref{cons2}), we thus
get
\begin{equation}\label{sch5}
   \left( -i\hbar\partial_t +\frac{p^2_x}{2m}+
   V\left(t, i\hbar(\partial_{p_x}-\theta
   \partial_t)\right)\right)\psi(t,p_x)=0.
\end{equation}
Note that contrary to what we had in the case of the basis  $\{|x,
p_t\rangle\}$ where the supplementary condition was independent of
time, here we have a time evolution equation. However, because of
the time derivative in the potential in ({\ref{sch5}) we may lose
the usual probability amplitude interpretation for $\psi(t,p_x)$ for
time derivatives of order higher than one, regardless of whether or
not the potential has an explicit dependence on time. It is
conceivable, nonetheless, that for certain forms of the potential a
probabilistic interpretation may be recovered by modifying the
product in the algebra of the wave functions or by redefining
hermicity, in analogy to what occurs in Feshbach-Villars formulation
of the Klein-Gordon
equation.\\

It is natural to ask how is (\ref{sch5}) related to (\ref{schro2})
for a time independent potential. For this purpose note that
\begin{equation}\label{pot}
\begin{split}
\langle t, p_x|V(\hat
x)|\psi\rangle=V\left(i\hbar(\partial_{p_x}-\theta
\partial_t)\right)\psi(t,p_x)= \hspace{2in}\\
\int dx\:dp_t
V\left(i\hbar(\partial_{p_x}-\theta
\partial_t)\right)\langle t,p_x|x,p_t \rangle\psi(x,p_t).
\end{split}
\end{equation}
But
\begin{equation}\label{ident}
\langle t,p_x|\hat x|x,p_t \rangle=x\langle t,p_x|x,p_t \rangle
=i\hbar(\partial_{p_x}-\theta\partial_t)\langle t,p_x|x,p_t \rangle.
\end{equation}
So
\begin{equation}\label{ident1}
V\left(i\hbar(\partial_{p_x}-\theta
   \partial_t)\right)\langle t,p_x|x,p_t \rangle=V(x)\langle t,p_x|x,p_t \rangle,
\end{equation}
and using
\begin{equation}\label{ident2}
\langle t,p_x|x,p_t \rangle=(2\pi\hbar)^{-1}
e^{-\frac{i}{\hbar}(xp_x -\theta p_t p_x -tp_t)},
\end{equation}
(see {\it e.g.} \cite{ros} for details of a procedure used to derive
a similar transition function), we get
\begin{equation}\label{ident3}
\langle t, p_x|V(\hat x)|\psi\rangle= (2\pi\hbar)^{-1}\int dx\:dp_t
V\left(x\right)e^{-\frac{i}{\hbar}(xp_x -\theta p_t p_x
-tp_t)}\psi(x,p_t).
\end{equation}
Finally, substituting this result in (\ref{sch5}) we get the
integro-differential equation
\begin{equation}\label{sch8}
\begin{split}
\left( -i\hbar\partial_t +\frac{p^2_x}{2m}\right)\psi(t,p_x)+\hspace{3in}\\
(2\pi\hbar)^{-2}\int dx\:dp_t \:dt^{\prime}\:dp_x^{\prime}
V\left(x\right)e^{-\frac{i}{\hbar}[x(p_x -p_x^{\prime}) -\theta p_t(
p_x -p_x^{\prime})
-(t-t^{\prime})p_t]}\psi(t^{\prime},p_x^{\prime})=0.
\end{split}
\end{equation}
On the other hand, if $\psi(x,p_t)$ is a solution of (\ref{schro2})
then
\begin{equation}\label{ident4}
\psi(t,p_x)=\int dx\:dp_t\: \langle
t,p_x|x,p_t\rangle\psi(x,p_t)=(2\pi\hbar)^{-1}\int dx\:dp_t\:
 e^{-\frac{i}{\hbar}(xp_x -\theta p_t p_x -tp_t)}\psi(x,p_t),
\end{equation}
is a solution of(\ref{sch5}). Indeed acting with $\left(
-i\hbar\partial_t +\frac{p^2_x}{2m}+
   V\left(t, i\hbar(\partial_{p_x}-\theta
   \partial_t)\right)\right)$ on (\ref{ident4}) and making use of (\ref{pot}) and (\ref{ident3})
we get
\begin{equation}\label{chanbasis}
\begin{split}
\left( -i\hbar\partial_t +\frac{p^2_x}{2m}+
   V\left(t, i\hbar(\partial_{p_x}-\theta
   \partial_t)\right)\right)\psi(t,p_x)=\hspace{2in}\\
(2\pi\hbar)^{-1}\int dx\:dp_t\:
 e^{-\frac{i}{\hbar}(xp_x -\theta p_t p_x -tp_t)}[p_t - \frac{\hbar^2}{2m}\partial_x^{2} +V(x)]\psi(x,p_t).
\end{split}
\end{equation}
Now, if $\psi(x,p_t)$ satisfies (\ref{schro2}) the right side of
(\ref{chanbasis}) is zero,
hence $\psi(t,p_x)$ as given by (\ref{ident4}) satisfies (\ref{sch5}). Q.E.D.\\

Let us now turn to the path integral quantization for this case and
the calculation of the propagator $\langle t(\tau_2),
p_x(\tau_2)|t(\tau_1), p_x(\tau_1)\rangle$. The appropriate solution
to the equations (\ref{paras}) for which $t, p_x$ are the fixed end
points of the action are
\begin{equation}\label{acb12}
    A_1  = p_t ,\;A_2  = 0 ,\;A_3  = 0,\;A_4  = \theta
p_t -x.
\end{equation}
Inserting this solution into the action (\ref{nonaction1}) we then
obtain,
\begin{equation}\label{act5}
S_2 = \int\limits_{\tau _1 }^{\tau _2 } {d\tau } \left( {p_t \dot t
+\theta p_t \dot p_x    -x \dot p_x - \lambda \varphi } \right),
\end{equation}
 Observe that the action (\ref{act5}) and the action
(\ref{act6}) are indeed related by a linear canonical transformation
generated by $F_1=p_x x -\theta p_t p_x - p_t t $.

The propagator for the admissible basis $\{|t, p_x\rangle\}$ is then
\begin{equation}\label{propa12}
\langle t(\tau_2), p_x(\tau_2)|t(\tau_1), p_x(\tau_1)\rangle = \int
{\mathcal D}t {\mathcal D}p_t {\mathcal D}x {\mathcal D}p_x
\delta(\chi)\delta(\varphi)\{\varphi,\chi\}^* \exp (\frac{i}{\hbar}
S_2),
\end{equation}
and for the boundary conditions that we are considering, the usual
gauge
\begin{equation}\label{gc01}
    t=f(\tau),
\end{equation}
is a good gauge condition.\\
Assuming now that the Hamiltonian is independent of $t$, we can
easily integrate (\ref{propa12}) over the variables $t$ and $p_t$,
using the gauge condition (\ref{gc01}) and the constraint, we get
\begin{equation}\label{propa22}
\begin{split}
\langle f(\tau_2), p_x(\tau_2)|f(\tau_1), p_x(\tau_1)\rangle = \hspace{2in}\\
\int{\mathcal D}x {\mathcal D}p_x (-1-\theta\partial_x V)\exp
\left(\frac{-i}{\hbar} \int_{f_1}^{f_2}df\left( (\theta
H+x)\frac{dp_x}{df} +H\right)\right),
\end{split}
\end{equation}
where the parametrization in the action has been eliminated. \\

Note that in the limit $\theta=0$ both (\ref{sch5}) and
(\ref{propa22}) reduce to the usual Quantum Mechanics. The same is
true for a free particle, as it is immediately evident from
(\ref{sch5}), and it is also follows  for (\ref{propa22}) since in
this case the Hamiltonian is independent of $x$, so by integrating
over this variable the term with
$\theta=0$ disappears. \\

\subsubsection{Basis $|p_t,p_x\rangle$} To conclude our analysis of the Dirac and
path integral
quantization realized on the three admissible bases for the extended
Heisenberg algebra (\ref{commu}) that we are studying in this
section, consider now the representation of the operators $(\hat t,
\hat x)$ in  $|p_t,p_x\rangle$. For this basis we have
\begin{equation}\label{ptpx}
   \hat t \psi(p_t, p_x)= (i\hbar \partial_{p_t}+a\theta p_x)\psi(p_t, p_x),\;\;
   \hat x \psi(p_t, p_x)= (i\hbar\partial_{p_x}+(1+a)\theta p_t)\psi(p_t,p_x).
\end{equation}
It is interesting to note that in this representation we have
introduced an extra parameter $a$, that can translate the
noncommutativity from the coordinate operator to the time operator.
(Observe that this characteristic is also present when we impose
noncommutativity of the space so we can also translate the
noncommutativity parameter from one coordinate to the another). For
this representation the constraint equation (\ref{qsup}) takes the
form
\begin{equation}\label{sch9}
   \left( p_t + \frac{p_x^2}{2m}
   +V\left(i\hbar \partial_{p_t}+a\theta p_x,i\hbar\partial_{p_x}+
   (1+a)\theta p_t\right)\right)\psi(p_t, p_x)=0.
\end{equation}
 Note that in this case, when the potential
is time independent so that (\ref{sch9}) reduces to
\begin{equation}\label{sch10}
   \left( p_t + \frac{p_x^2}{2m}
   +V\left(i\hbar \partial_{p_x}+(1+a)\theta p_t \right) \right)\psi(p_t, p_x)=0,
\end{equation}
we do have noncommutative corrections except when we choose the
parameter $a=0$, or for the case of a free particle.\\
For the path integral formulation in this basis, an appropriate
action (having $p_t, p_x$ as fixed end-points) is given by
\begin{equation}\label{act7}
S_3 = \int\limits_{\tau _1 }^{\tau _2 } d\tau \left( -t \dot p_t
+a\theta p_t \dot p_x -(1-a)\theta p_x \dot p_t    -x \dot p_x -
\lambda \varphi ) \right),
\end{equation}
from which we can obtain results  equivalent to those derived from
the analysis of the constraint equation (\ref{sch9}).\\

Contrary to the actions $S_1$ and $S_2$ which are unique
solutions of (\ref{symstr}) for their corresponding fixed
end-points, there are several canonically equivalent admissible
actions with fixed points $p_t, p_x$. Thus, for example, $S_4
=\int_{\tau_1}^{\tau_2} \:d\tau(-t\dot p_t -\theta p_x \dot p_t -x\dot p_x)$
can be obtained from $S_3$ by substracting the total derivative of $
F_2 =a\theta p_t p_x$ from the integrand in $S_3$. Other canonically
equivalent actions follow from $S_3$ and $S_4$ by means of the
generator $F_3= \theta p_t p_x$. \\

\subsubsection{Noncanonical related actions}  Up to this point we have
considered path integral quantizations based on actions which are
compatible with the extended Heisenberg algebra (\ref{commu}),
derived by means of the Dirac quantization procedure. There are,
however, other solutions to the equations (\ref{paras}) which,
although indistinguishable at the classical level from the ones
considered so far, they are not canonically related to them, in the
sense that there is no generating function for mapping canonically
the actions resulting from these solutions to the ones previously
considered. We shall see that in these cases the transformations
needed for fixing the end-points required for a path integral
quantization are actually transformations which map the original
phase-space variables with symplectic structure (\ref{commu}) to
another set of variables related to the canonical symplectic
structure (\ref{omega}). Classically, as it is well known from the
Darboux theorem \cite{Arnold}, this map is always possible (at least
locally). To each of these Darboux maps corresponds, however,
 a different quantum mechanics, generated by
what in some works in the literature has been called the equivalent
of the Seiberg-Witten map for ``noncommutative quantum
mechanics".\\

To exhibit in more detail the above considerations, let us begin
with the solutions:
\begin{equation}\label{solas}
\begin{gathered}
  A_1  = p_t ,\;A_2  = p_x ,\;A_3  = 0,\;A_4  = \theta
p_t , \hfill \\
  A_1  = p_t ,\;A_2  = p_x ,\;A_3  =  - \frac{\theta }
{2}p_x ,\;A_4  = \frac{\theta }
{2}p_t . \hfill \\
\end{gathered}
\end{equation}

With the first set of equations in (\ref{solas}), the canonical
action takes the form
\begin{equation}\label{act1}
S_5 = \int\limits_{\tau _1 }^{\tau _2 } {d\tau } \left( {p_t (t +
\theta p_x )^ \bullet   + p_x \dot x - \lambda \left( {p_t  +
H(t,x,p_x)} \right)} \right).
\end{equation}
We therefore see from (\ref{act1}) that from the original
phase-space variables of the theory we do not have a set of fixed
end-points for the action from which a quantization can be
developed. Nonetheless a natural pair $ (\tilde t,x)$ can be
constructed by making the change of variables
\begin{equation}\label{ttilde}
\tilde t: = t + \theta p_x, \;\; \tilde x=x,
\end{equation}
where $\tilde t$ is a new canonical variable associated to the time.
In terms of this new pair of variables, the symplectic structure is
reduced to (\ref{omega}), and
%i.e., this new time is commutative, it satisfies
introducing this new time in the action (\ref{act1}),
 results in
\begin{equation}\label{act2}
S_5 = \int\limits_{\tau _1 }^{\tau _2 } {d\tau } \left( {p_t \dot
{\tilde t} + p_x \dot x - \lambda \left( {p_t  + H(\tilde t - \theta
p_x ,x,p_x )} \right)} \right).
\end{equation}
Note that if the original Hamiltonian was time-dependent, the
modified one introduces a new  kind of interaction that is
proportional to the parameter $\theta$ of noncommutativity and to
the momenta in the spatial direction. Also note that in terms of the
modified symplectic structure (\ref{omega}) the Dirac brackets
(\ref{diracbraz}) lead, upon quantization, to the commutators
\begin{equation}\label{comttilde}
\left[ {\tilde t,p_t } \right] = i\hbar,\ \ \left[ {\tilde t,x}
\right] = 0, \ \ \left[x,p_x \right]=i\hbar,\ \
\left[x,p_t\right]=0,\ \ \left[p_x,p_t\right]=0.
\end{equation}
From these commutators we clearly see that a new complete set of
commuting observables is  $(\hat{\tilde t}, \hat x)$, which label
the admissible associated basis of coordinate states $\{\left|
{\tilde t,x} \right\rangle\}$.
The Dirac's supplementary condition in this basis is now,
\begin{equation}\label{sch1}
\left( { - i\hbar\frac{\partial } {{\partial \tilde t}} + \hat
H(\tilde t +i\hbar \theta \partial_x ,x,-i\hbar\partial_x )}
\right)\psi (x,\tilde t) = 0,
\end{equation}
and we note that in the case that the Hamiltonian does not depend
explicitly on the time the Schr\"odinger equation is not modified
by the noncommutativity.\\

Now, if we consider the second set in (\ref{solas}) of solutions to
(\ref{paras}) the resulting action is given by
\begin{equation}\label{act3}
S_6 = \int\limits_{\tau _1 }^{\tau _2 } {d\tau } \left( {p_t (t +
\frac{\theta } {2}p_x )^ \bullet   + p_x (x - \frac{\theta } {2}p_t
)^ \bullet   - \lambda \left( {p_t  + H(t,x,p_x)} \right)} \right).
\end{equation}
Following the same logic as in the previous case, it is natural to
introduce in this equation  the new set  ($\check{t}=t +
\frac{\theta } {2}p_x  , \check{x}=x - \frac{\theta } {2}p_t$) of
time and spatial coordinate. Here then the action (\ref{act3}) is
reduced to
\begin{equation}\label{act4}
S_6 = \int\limits_{\tau _1 }^{\tau _2 } {d\tau } \left( {p_t \dot
{\check{t}} + p_x \dot{\check x} - \lambda \left( {p_t + H(\check t
-\frac{\theta }{2}p_x,\check x+\frac{\theta }{2}p_t,p_x)} \right)},
\right)
\end{equation}
and, upon Dirac quantization, the corresponding new set of dynamical
observables satisfies the following commutation relations,
\begin{equation}\label{commtx}
\left[\hat {\check{t}} ,\hat {\check{x}}  \right] = 0, \ \
\left[\hat {\check{t}} ,p_t \right] = i\hbar, \ \ \left[\hat
{\check{x}} ,\hat p_x  \right] = i\hbar, \ \ \left[\hat p_t,\hat
p_x\right]=0.
\end{equation}
Using as a complete set of commuting observables the variables
$(\hat {\check{t}},\hat {\check{x}})$, the new supplementary Dirac
condition is
\begin{equation}\label{sch2}
\left(  - i\hbar\frac{\partial }{\partial \check{t}} + H(\check{t} +
i\frac{\hbar\theta }{2}\partial_{\check{x}} ,\; \check{x} -i
\frac{\hbar\theta }
{2}\partial_{\check{t}},\;-i\hbar\partial_{\check{x}} ) \right)\psi
(\check{t} ,\check{x} ) = 0.
\end{equation}
For this Schr\"odinger equation we see that including the case when
the Hamiltonian does not depend explicitly on time we do have
modifications originated by the noncommutativity. Furthermore we see
that the new theory could be non-unitary, since partials with
respect to $\check{t}$ appear to an order that depends on the kind
of interaction. This type of quantization can been formulated
directly by using the Moyal product:
\begin{equation}\label{moyalprod}
H(\check{t} + i\frac{\hbar\theta } {2}\partial_{\check{x}} ,\;
\check{x} -i \frac{\hbar\theta }
{2}\partial_{\check{t}},\;-i\hbar\partial_{\check{x}} )\psi
(\check{t} ,\check{x} ) =H(\check{t},\; \check{x}
,\;-i\hbar\partial_{\check{x}} )\star_{\theta}\psi (\check{t}
,\check{x} ),
\end{equation}
where
\begin{equation}\label{moyal}
\star_{\theta}=
\exp\left[i\frac{\hbar\theta}{2}\left(\overleftarrow\partial_{\check{t}}
\overrightarrow\partial_{\check{x}}-\overleftarrow\partial_{\check{x}}
\overrightarrow\partial_{\check{t}}\right)\right].
\end{equation}
So for this selection of symplectic potentials the theory is not
unitary and this result is equivalent to the obtained in Ref.
\cite{Chaichian} in the context of noncommutative field theory.

To quantize these two cases by means of the path integral method we
make use of the basis $\{\left| {\tilde t,x} \right\rangle\}$ and
the respective actions (\ref{act2}) and (\ref{act4}) to compute the
propagator
\begin{equation}\label{propa33}
    \langle \tilde t_2, x_2 | \tilde t_1, x_1 \rangle.
\end{equation}
Following the normal procedure to quantize a theory with first class
constraints \cite{Henneaux}, we have only two extra points to
consider. First we have to impose a gauge condition, which in this
case can be the normal canonical gauge  $\tilde t=f(\tau)$, since in
difference with the approach used in \cite{Pinzul} and
\cite{Banerjee} we are imposing the noncommutativity at the level of
the action, using the symplectic structure, and not at the level of
the gauge condition. The second point that we need to take into
account is the extra appearance in the Hamiltonian of the $\theta
p_x$ shifted term when we have a $t$ dependent theory,
 this can imply that it may not be possible to compute the path
integral over the momenta. These are however the usual problems that
one finds when computing path integrals with actions in terms of
variables with powers larger than two.\\

One additional point to notice is that for both types of solutions
of the equations (\ref{paras}) considered in this section, the Dirac
constraint is not modified, since in both cases the new time is
canonical conjugated to the original $p_t$ and then the constraint
generates the parametrization invariance. It is not difficult to see
that this is not the case when the above analysis is extended to the
more general case of symplectic structures that upon quantization
result in an extended Heisenberg algebra that includes
noncommutativity of the momenta. \\
For such a generalization one would have to consider a symplectic
structure of the form
\begin{equation}\label{oab12}
\omega ^{ab}  = \left( {\begin{array}{*{20}c}
   0 & \theta  & 1 & 0  \\
   { - \theta } & 0 & 0 & 1  \\
   { - 1} & 0 & 0 & \beta  \\
   0 & { - 1} & -\beta & 0  \\
\end{array} } \right),  \ \ \
\omega _{ab}  = \frac{1}{\gamma}\left( {\begin{array}{*{20}c}
   0 & \beta & -1 & 0  \\
   -\beta & 0 & 0 & -1  \\
   1 & 0 & 0 & \theta   \\
   0 & 1 & -\theta & 0  \\
\end{array} } \right),
\end{equation}
where
\begin{equation}\label{paramet}
\gamma  =  1 - \beta \theta .
\end{equation}
Here the quantization of the Dirac brackets would then result in the
extended Heisenberg algebra
\begin{equation}\label{commu2}
[\hat t,\hat x] = i\hbar\theta, \ \ [\hat x,\hat p_x] = i\hbar,\ \
[\hat t,\hat p_t] = i\hbar ,\ \ [\hat p_t,\hat p_x] = i\hbar\beta,
\end{equation}
and, in contradistinction to what occurred for the previously
considered symplectic structure, we would only have two complete
sets of commuting fundamental observables: $(\hat x, \hat p_t) $ and
$(\hat t, \hat p_x)$,
with their respective admissible bases: $\{|x,p_t \rangle\}$ and $\{|t,p_x \rangle\}$.\\
Except for some differences such as the ones mentioned above, the
analysis of the Dirac and path integral quantizations relative to
these bases, as well as others resulting from considering canonical
transformations of their respective associated actions followed by
Darboux maps, is qualitatively similar (see \cite{nos2}) to what we
have already done, so for the sake of brevity we shall omit the
details here.

 Rather, and in preparation for a future investigation of how our analysis of
space-time noncommutativity in the discrete realm of quantum
mechanics can be extended to the continuum of relativistic field
theory, we turn next our consideration to the case of a relativistic
particle.

\subsection{Space-time Noncommutativity for a Relativistic particle}
Our starting point is the action for the free relativistic particle
\begin{equation}\label{relaact}
S = \int\limits_{\tau _1 }^{\tau _2 } {d\tau } \left( { - m\sqrt { -
\eta _{\alpha \beta } \dot x^\alpha  \dot x^\beta  } } \right), \ \
\ \ \alpha=1,\dots,n.
\end{equation}
In Hamiltonian form we have
\begin{equation}\label{relaham}
S = \int\limits_{\tau _1 }^{\tau _2 } {d\tau } \left( {p_\alpha \dot
x^\alpha   - \lambda \left( {p^2  + m^2 } \right)} \right),
\end{equation}
where now the first class primary constraint $\varphi$ is given by
\begin{equation}\label{relacon}
\varphi  = p^2  + m^2  \approx 0.
\end{equation}
As discussed above in Sec. 3, for an arbitrary symplectic structure
the
 action (\ref{relaham}) has the
form
\begin{equation}\label{actr1}
S = \int\limits_{\tau _1 }^{\tau _2 } {d\tau } \left( {A_a (z)\dot
z^a  - \lambda \left( {\varphi (z)} \right)} \right),  \ \ \
z^a=(x^\alpha,p_\alpha), \ \ a=1,\dots,2n.
\end{equation}
Again, arising from the definition of the momenta, we have the
primary constraints
\begin{equation}\label{primaconrel}
\chi _a  = {p_z}_a  - A_a \left( z \right).
\end{equation}
These constraints are second class and the corresponding Dirac
brackets are identical in form to those in the  non-relativistic
case,
given by Eq.(\ref{diracbraz}). \\

Let us consider now a symplectic structure which is determined by
the following Dirac brackets involving the space-time and momentum
variables:
\begin{equation}\label{diracbr}
\{ x^\alpha  ,x^\beta  \}^\ast = \theta ^{\alpha \beta }, \ \ \{
x^\alpha  ,p_\beta   \}^\ast = \delta^\alpha_\beta,
\end{equation}
where $\theta ^{\alpha \beta }$ is a constant antisymmetric tensor.
Then, the symplectic structure takes the explicit form:
\begin{equation}\label{relasymple}
\omega ^{ab}  = \left( {\begin{array}{*{20}c}
   {\theta ^{\alpha \beta } } &   \mathbb{I} \\
   -\mathbb{I} & 0  \
\end{array} } \right), \  \ \ \
\omega _{ab}  = \left( {\begin{array}{*{20}c}
   0 & -\mathbb{I}  \\
   \mathbb{I} & {\theta ^{\alpha \beta } }  \\
\end{array} } \right).
\end{equation}
Note, as it was the case before, that the solutions of
\begin{equation}\label{relsym}
\omega _{ab}  = \partial _a A_b  - \partial _b A_a,
\end{equation}
for the generating potentials $A_a$, are not unique, but they are
all related by canonical transformations. One possible covariant
solution is
\begin{equation}\label{apotre3}
A_\alpha   =0 ,\ \ \ A_{n + \alpha }  = -x_\alpha -
\frac{{\theta_{\alpha \beta } }} {2}p^\beta,\ \ \alpha=1\dots n.
\end{equation}
Introducing this symplectic potential in the action (\ref{actr1}),
we obtain
\begin{equation}\label{actr31}
S = \int\limits_{\tau _1 }^{\tau _2 } {d\tau } \left(- x^\alpha \dot
p_\alpha + \frac{\theta^{\alpha \beta } } {2}p_\alpha  \dot p_\beta
   - \lambda \left( {p^2  + m^2 } \right) \right).
\end{equation}
So the variables with fixed end points in the action are the momenta
$p_\alpha$.  Dirac quantization in this case results in the
commutators
\begin{equation}\label{hei2}
[\hat x^\alpha  ,\hat x^\beta ] =i\hbar \theta ^{\alpha \beta }, \ \
[\hat x^\alpha , \hat p_\beta]=i\hbar\delta^\alpha_\beta, \ \ [\hat
p_\alpha , \hat p_\beta]=0,
\end{equation}
and the supplementary condition
\begin{equation}\label{dsc}
\hat \varphi(\hat x, \hat p)\psi(p)=(p_{\alpha}p^{\alpha} +m^2)
\psi( p)=0,
\end{equation}
referred to the admissible basis $\{|p\rangle\}$.
As is to be
expected, this merely states that $(p_{\alpha}p^{\alpha} +m^2)=0$.\\

To compute the propagator for the theory (\ref{actr31}) using path
integrals, the more convenient technique is to use a non-canonical
gauge and the BFV-BRST path integral procedure \cite{Henneaux}. The
full action, after introducing the gauge fixing term and ghost
terms, is
\begin{equation}\label{actbrst}
S = \int\limits_{\tau _1 }^{\tau _2 } {d\tau } \left(-( x^\alpha +
\frac{\theta^{\alpha \beta } } {2}p_\beta ) \dot p_\alpha -
   \lambda\dot \pi -  P\dot {\bar C} + \dot C \bar P -i\bar P P -
\lambda \left( {p^2  + m^2 } \right) \right).
\end{equation}
Here the boundary conditions on the ghost, the momenta conjugate to
the coordinates $p_\alpha$ and the Lagrange multiplier $\pi$ are
\begin{eqnarray}\label{bcrel}
&&\pi(\tau_1)=\pi(\tau_2)=\bar C(\tau_1)=\bar
C(\tau_2)=C(\tau_1)=C(\tau_2)=0, \\
&& p_\alpha(\tau_1)=p_{\alpha 1}, \ \ \ p_\alpha(\tau_2)=p_{\alpha
2}.
\end{eqnarray}
From the path integral over the ghosts we get a multiplicative
factor of $(\tau_1 -\tau_2)$, this term is very useful since it
allows to eliminate the dependence of the propagator on the
parameter $\tau$.  Using the path integral over $x^\alpha$, we
obtain delta functions that we then use to integrate over the
momenta $p_\alpha$. As a result these integrals cancel the
$\theta^{\alpha\beta}$ correction term, since this term is
multiplied by $\dot p_\alpha$. So, finally we get the usual
propagator in the basis where the momenta are fixed at the end
points
\begin{equation}\label{relp}
\langle p_\beta(\tau_2)  | p_\alpha(\tau_1) \rangle =
\frac{-i\eta_{\alpha\beta}\delta(p_{\alpha 2}-p_{\alpha
1})}{p^2+m^2}.
\end{equation}
This result is fully consistent with the previous result
(\ref{dsc}).

Other admissible bases compatible with the Heisenberg algebra
(\ref{hei2}) are obtained from (\ref{actr31}) by a canonical
transformation generated by $F=p_\alpha x^\alpha$, for $\alpha$
fixed. These sets of admissible
bases are \\
$\{|x^\alpha, p_\beta, p_\gamma, p_\lambda\rangle
;\;\alpha\neq\beta\neq\gamma\neq\lambda\}$. Refered to them, the
Dirac subsidiary condition results in
\begin{equation}\label{other}
\hat \varphi(\hat x, \hat p)\psi(x^\alpha, p_\beta, p_\gamma,
p_\lambda)= \left ( -\hbar^2 (\partial_{x^a})^2 +(p_\beta)^2
+(p_\gamma)^2 +(p_\lambda)^{2} \right) \psi(x^\alpha, p_\beta,
p_\gamma, p_\lambda)=0,
\end{equation}
where indices here are not summed over.\\
So, even though the deformation parameter $\theta$ does not appear
in these constraint equations the space-time noncommutativity is
reflected in
their violation of Lorentz invariance.\\
On the other hand, canonically transforming (\ref{actr31}) with
$F=p_\alpha x^\alpha$, where now we sum over $\alpha$, we get, after
regrouping terms,
\begin{equation}\label{actr3}
S = \int\limits_{\tau _1 }^{\tau _2 } {d\tau } \left( {p_\alpha
(x^\alpha   + \frac{{\theta ^{\alpha \beta } }} {2}p_\beta  )^
\bullet   - \lambda \left( {p^2  + m^2 } \right)} \right).
\end{equation}
Here we see that it is natural to define as fixed end-point
variables of the action the new set of coordinates given by
\begin{equation}\label{xtilde1}
\tilde x^\alpha   = x^\alpha   +\frac{{\theta ^{\alpha \beta } }}
{2}p_\beta.
\end{equation}
The Dirac bracket between these new coordinates vanishes and, in
consequence, so does  their commutator:
\begin{equation}\label{relxcom}
\left[ {\tilde x^\alpha  ,\tilde x^\beta  } \right] = 0,
\end{equation}
while
\begin{equation}\label{relxcom2}
\left[ {\tilde x^\alpha  , p_\beta  } \right] =
i\hbar\delta^\alpha_\beta.
\end{equation}
Note, however, that (\ref{xtilde1}) is a Darboux map and not a
canonical transformation of the action (\ref{actr31}). Consequently
this is a different Dirac quantization, related to the canonical
symplectic form and not to the original one given by
(\ref{relasymple}). The Dirac supplementary condition in this case
is
\begin{equation}
\hat \varphi(\hat {\tilde x}, \hat p)\psi(\tilde x)=
\left(-\partial_{\tilde x_\alpha}\partial^{\tilde x_\alpha} + m^2
\right)\psi(\tilde x)=0.
\end{equation}
So, quantizing the theory in this way we obtain that a relativistic
particle satisfies the Klein-Gordon equation, and thus arrive at the
well known result that for a free particle we do not obtain any
deformation of the theory. However, if we consider that the particle
lives in a given background, we will get the deformation produced by
the new choice of coordinates.\\
To further illustrate this point, consider the interaction of the
relativistic particle with a constant external field. Here the
constraint will be of the form
\begin{equation}\label{cons5}
    \left(\Pi_\mu - \frac{1}{2} F_{\mu\nu}x^\nu\right)
    \left(\Pi^\mu - \frac{1}{2} F^{\mu\sigma}x_\sigma\right)+m^2\approx
    0.
\end{equation}
Using the $\tilde x^\alpha$ coordinates, which will have the same
form as in (\ref{xtilde1}), except for the substitution $p_\beta \to
\Pi_\beta$, the Dirac supplementary condition in the basis
$\{|\tilde x^\alpha\rangle\}$ is of the form
\begin{equation}\label{k-g2}
\begin{split}
\left[\left(-i\hbar\partial_\mu - \frac{1}{2} F_{\mu\nu}\left(\tilde
x^\nu+i\hbar\frac{1}{2}\theta^{\nu\rho}\partial_\rho\right)\right)\right.\times\hspace{1in}\\
  \left.  \left(-i\hbar\partial^\mu - \frac{1}{2}
    F^{\mu\sigma}\left(\tilde x_\sigma +i\hbar\frac{1}{2}
    \theta_{\sigma\rho}\partial^\rho\right)\right)+m^2 \right]\psi(\tilde x)=
    0,
\end{split}
\end{equation}
which indeed shows corrections containing the deformation  parameter
$\theta$.

\section{Concluding remarks}

We have seen that according to the Dirac quantization scheme for
constrained systems, it is the first class constraints and the
symplectic structure resulting from the Dirac brackets that uniquely
define a particular quantum theory, irrespectively of the fact that
there are many possible solutions for the potentials $A_a$
corresponding to the same symplectic structure $\omega$. On the
other hand, if we use these solutions as the starting point for
evaluating the action in the path integral formulation, then
depending on the type of solutions that we propose for the equations
(\ref{symstr}), we could get different quantizations. We have seen
moreover, that if there is a linear canonical transformation
relating these actions, as is the case for the actions $S_1$, $S_2$
and $S_3$ considered in subsections 4.1.1-4.1.3, then the
corresponding quantizations are actually equivalent to each other
and differ only by the fact that they are referred to the three
admissible bases compatible with the extended Heisenberg algebra
(\ref{commu}). Indeed, the phases of the quantum mechanical
transition functions corresponding to changes between these bases
({\it cf. e.g.} Eq. (\ref{ident2})) are nothing other than the
classical generating functions of the linear canonical
transformations among the three actions, and the associated
symplectic transformation leaving invariant their common symplectic
structure $\omega$ is, for each of these three cases,
the identity element of the group.\\

Alternatively, for the type of solutions to (\ref{symstr}) leading
to the actions considered in subsection 4.1.4, the situation is
actually quite different because there is no generating function
that permits to canonically transform such actions to the ones
previously considered, and because at the classical level fixing the
end-points of these actions involves a change of variables in
extended phase-space which results in a Darboux map from the
original symplectic structure to the canonical one given by (\ref{omega}).\\
Quantizing in these cases via either the Dirac or path
integral formalisms is then tantamount to applying standard quantum
mechanics with a Hamiltonian modified with the new variables, which
are formally promoted to the rank of operators satisfying the
commutation relations (\ref{commtx}). But in axiomatic quantum
mechanics the operators acting on vectors in Hilbert space are
observables, {\it i.e.} operators functions of the basic dynamical
variables of the theory, with eigenvalues given by quantities
measurable by experiment. For the systems we have been considering
and the construction followed in subsection 4.1.4, this would imply
that the new time and coordinate variables are the observables of
the theory and, since they obey the commutation relations
(\ref{commtx}), the new time and coordinate operators commute.
Physically this would then mean that experiments could be designed
to measure simultaneously the eigenvalues of these space-time
operators. This, however, begs the question of what is then the true
physical interpretation for the $\theta$ parameter that appears in
the modified quantum expressions of the theory, such as the
Hamiltonian? We could try to further argue that both the old and new
space-time operators are observables and that $\theta$ reflects the
noncommutativity of the old observables. This, however, brings in a
somewhat Bohmian flavor of hidden variables to the new quantization
which is, to say the least, subject to questioning (for additional
arguments regarding this issue see \cite{barb}). Thus, from our
point of view, it would seem preferable to conclude that in the case
of the quantizations discussed in subsection 4.1.4, the term
``space-time noncommutativity" is a misnomer. Nonetheless, since the
different quantizations here discussed lead to different (at least
conceptually) experimental predictions, it is experiment then that
will determine which, if any, of these
theories can be closer related to reality. \\
The same can be said regarding the different cases discussed in
Section 4.2
for the relativistic particle.\\

Of course it could also be contended that the use of the Dirac and
path integral quantizations, which have been so successful in
extending classical mechanics and field theory to a certain range of
the quantum realm, is not justified {\it a priori } when dealing
with distances of the order of the Planck length where quantum
gravity becomes relevant. This could very well be so and it may
involve having to drop the very concept of manifold, which underlies the
mathematics of all of our present day physical constructions, in
favor of new geometrical paradigms in which quantization is built in
{\it ab initio}, such as the noncommutative geometry proposed by
Connes \cite{connes} a few years ago. Be it as it may, we believe
that the analysis presented here, the more axiomatic one
presented in \cite{ros} and references within,
as well as many other related works that have appeared in the
literature, could provide some guidance for further work in that
ultimate direction.

\bibliographystyle{amsalpha}

\end{document}